\documentclass[prb,aps,twocolumn,superscriptaddress,floatfix,showpacs]{revtex4-1}
\usepackage{amssymb,amsmath,bm,graphicx,textcomp}

\begin{document}

\title{Order from structural disorder in $XY$ pyrochlore antiferromagnet
$\rm Er_2Ti_2O_7$}

\author{V. S. Maryasin}
\affiliation{
University Grenoble-Alps and INAC-SPSMS, F-38000 Grenoble, France}
\author{M. E. Zhitomirsky}
\affiliation{Service de Physique Statistique, Magn\'etisme et Supraconductivit\'e,
UMR-E9001 CEA-INAC/UJF, 17 rue des Martyrs, 38054 Grenoble Cedex 9, France}

\begin{abstract}
Effect of structural disorder is investigated for an $XY$ pyrochlore
antiferromagnet with continuous degeneracy of classical ground states. Two types of
disorder, vacancies and weakly fluctuating exchange bonds, lift degeneracy
selecting  the same subset of classical ground states. Analytic and numerical results
demonstrate that such an ``order by structural disorder'' mechanism competes with
the effect of thermal and quantum fluctuations. Our theory predicts that a small
amount of nonmagnetic impurities in $\rm{Er_2Ti_2O_7}$ will stabilize the coplanar
$\psi_3$ ($m_{x^2-y^2}$) magnetic structure as opposed to the $\psi_2$ ($m_{3z^2-r^2}$)
state found in pure material.
\end{abstract}
\date{\today}
\maketitle
\maketitle

\section{Introduction}

Geometrically frustrated magnets with competing exchange interactions often display continuous,
symmetry unrelated degeneracy of classical ground states. Such an `accidental' degeneracy
may be lifted by weak additional interactions. Those are always present in real materials
but can significantly vary even between similar compounds. Therefore, a lot of
studies on frustrated magnets have been devoted to understanding
the universal degeneracy-lifting  mechanism produced by thermal and quantum
fluctuations. The corresponding concept  named ``order by disorder'' was pioneered by Villain
\textit{et al.} [\onlinecite{Villain80}] and Shender [\onlinecite{Shender82}] and, in a nutshell,
relates the ground-state selection to softer excitation spectrum for certain degenerate states.
Being investigated for numerous spin models, the order
by disorder mechanism finds so far only a few realizations in
magnetic materials. Perhaps the clearest examples of the order by disorder selection
are provided by the 1/3-magnetization plateau in triangular-lattice antiferromagnets
[\onlinecite{Inami96,Smirnov07,Fortune09,Susuki13}] and by a zero-field noncoplanar spin structure
of  the $XY$ pyrochlore antiferromagnet $\rm Er_2Ti_2O_7$ [\onlinecite{Zhitomirsky12}, \onlinecite{Savary12}].

Weak lattice disorder, if present in a magnetic solid, changes locally parameters of the spin Hamiltonian
and can also affect the ground state selection [\onlinecite{Henley89,Fyodorov91,Weber12,Maryasin13}].
For a few studied models, the structural disorder tends to select classical
ground-states in precisely the opposite manner compared to the thermal and quantum
effects. These include an orthogonal magnetic structure for the $J_1$--$J_2$
square lattice antiferromagnet [\onlinecite{Henley89}, \onlinecite{Weber12}] and a conical
state for the Heisenberg triangular-lattice antiferromagnet in an external field [\onlinecite{Maryasin13}].
In our previous work, such a difference was explained by opposite signs of
effective biquadratic interactions generated by two types of the order from disorder:
thermal and quantum fluctuations yield a negative biquadratic exchange [\onlinecite{Heinila93}, \onlinecite{Canals04}],
whereas bond and site disorder produce a positive biquadratic term [\onlinecite{Maryasin13}].
Note, that in a rather different context a positive biquadratic coupling in ferromagnetic
multilayers was attributed to interface roughness [\onlinecite{Slonczewski91}].

The known examples bring up a problem of further generalization of the above principle.
Specifically, for some highly symmetric frustrated spin models, an effective interaction
that is able to lift the classical degeneracy appears only beyond the fourth-order
terms in the Landau energy functional. Then, a biquadratic coupling of either
sign leaves degeneracy intact raising again the question about the outcome of the order from
structural disorder selection.

\begin{figure}[t]
\centerline{
\includegraphics[width=0.95\columnwidth]{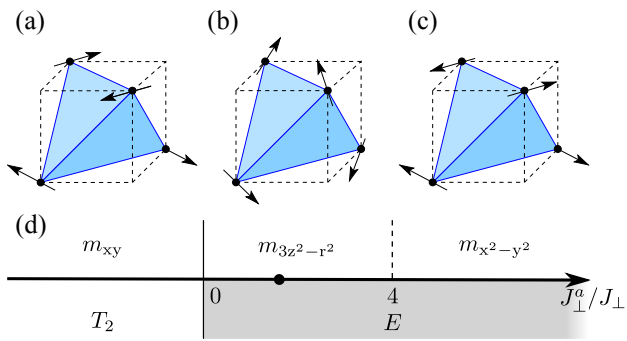} \hspace{2mm}
}
\caption{
(Color online) Ordered magnetic structures of the $XY$ pyrochlore antiferromagnet:
(a) the coplanar Palmer-Chalker state $m_{xy}$, (b) the noncoplanar state $m_{3z^2 - r^2}$,
and (c) the coplanar state $m_{x^2-y^2}$.
(d) Schematic phase diagram of the model (\ref{Ham1}) showing the lowest-energy
magnetic structures depending on the ratio $J_{\perp}^a / J_{\perp}$.
Circle  marks the value of $J_{\perp}^a/J_{\perp}$ in $\rm{Er_2Ti_2O_7}$.
}
\label{fig:states}
\end{figure}

As a matter of fact, such a highly symmetric frustrated spin model is realized in the anisotropic $XY$ pyrochlore
$\rm Er_2Ti_2O_7$ [\onlinecite{Champion03,Poole07,Ruff08,Sosin10,Dalmas12,Bonville13,Ross14}].
This pyrochlore material orders at $T_N\simeq 1.2$~K into a noncoplanar
$k=0$ antiferromagnetic structure called the $\psi_2$ state  [\onlinecite{Champion03}, \onlinecite{Poole07}].
To emphasize its symmetry properties we shall denote this state as $m_{3z^2-r^2}$ in the following.
At the mean-field level, there is no energy difference between the $m_{3z^2 - r^2}$ state
and the coplanar $m_{x^2 - y^2}$ ($\psi_3$) state, see Fig.~\ref{fig:states}.
The two states form a basis of the $E$ irreducible representation of the tetrahedral
point group and can be continuously turned into each other by simultaneous rotation
of four sublattices. Such degeneracy persists even with an extra biquadratic exchange,
but the harmonic spin-wave calculations indicate that fluctuations choose
the noncoplanar $m_{3z^2 - r^2}$ state
[\onlinecite{Champion04,Zhitomirsky12,Savary12,Wong13,McClarty14,Zhitomirsky14}].
Therefore, it is instructive to study the role of structural disorder, in particular,
nonmagnetic vacancies, on degeneracy lifting for the $XY$ pyrochlore antiferromagnet. This is
even more so in view of well established experimental possibility to systematically substitute
nonmagnetic ions  in pyrochlore materials [\onlinecite{Ke07}, \onlinecite{Chang10}].

In the present work we study theoretically the effect of structural disorder on the
classical $XY$ pyrochlore antiferromagnet. Similar to other frustrated models,
structural disorder favors in this case a different subset of classical ground states
compared to those selected by quantum and thermal fluctuations. Specifically, we predict
that nonmagnetic impurities substituted into $\rm{Er_2Ti_2O_7}$ will stabilize
the coplanar $m_{x^2-y^2}$ antiferromagnetic state. The paper is
organized as follows. Section II describes the spin model appropriate for anisotropic
$XY$ pyrochlores. In Sec.~III we develop an analytical approach to the problem
of the ground state selection in the framework of the real-space perturbation theory.
Section~IV contains our main analytic result: corrections to the classical ground
state energy produced by weak bond and site disorder. Numerical results in support of
the analytic analysis include the ground-state energy minimization and Monte Carlo
simulations and are described in Sec.~V. In Sec.~VI we discuss competition
between state selection produced by quantum fluctuations and the structural disorder
in view of possible realization in $\rm{Er_2Ti_2O_7}$. Finally, Sec.~VII contains 
conclusions and gives further outlook.

\section{Spin Model}
\label{sec:Model}

Low-temperature magnetic properties of $\rm{Er_2Ti_2O_7}$ and a number of other insulating
pyrochlore materials are well approximated by an effective pseudo-spin-1/2  model for
interacting Kramers doublets produced by strong crystal-field splitting. In the case of
$\rm{Er_2Ti_2O_7}$, the crystalline electric field determines the predominantly planar
character of the lowest-energy Kramers doublets [\onlinecite{Champion03}]. Correspondingly,
the effective spin-1/2 Hamiltonian features the anisotropic $XY$ interactions:
[\onlinecite{Zhitomirsky12}]
\begin{eqnarray}
\hat{\cal H} =  \sum_{\langle ij\rangle} \bigl[ J_{\perp} {\bf S}^{\perp}_{i}\cdot
{\bf S}^{\perp}_{j} + J_{\perp}^{a} ({\bf S}^{\perp}_{i}\cdot \hat{\bf r}_{ij})
({\bf S}^{\perp}_{j}\cdot \hat{\bf r}_{ij}) \bigr]\,.
\label{Ham1}
\end{eqnarray}
Here $\hat{\bf r}_{ij} = ({\bf r}_{i} - {\bf r}_{j}) / |{\bf r}_{i} - {\bf r}_{j}|$
is a unit vector in the bond direction and spin components are taken with respect to the local axes
such that the $z_i$ direction coincides with the  $[111]$ axis and
$\mathbf{S}^{\perp}_i$ refers to the projection onto the orthogonal plane.
We are interested in the case of antiferromagnetic
exchange interactions $J_{\perp},J_{\perp}^a > 0$ relevant to $\rm{Er_2Ti_2O_7}$
and assume an arbitrary value of spin $S$ in order to separate classical and quantum effects
in the framework of the semiclassical $1/S$ expansion. Further details on geometry of
a pyrochlore lattice and different forms of the spin Hamiltonian (\ref{Ham1}) are provided
in Appendix~\ref{sec:AppSGeom}.

We begin with description of the classical ground states of the spin model (\ref{Ham1}).
Depending on the sign of $J_{\perp}^a$, magnetically ordered states
belong to one of the two different classes, which transform according to $E$ $(\Gamma_5)$
or $T_2$ $(\Gamma_7)$ irreducible representations of the tetrahedral point group.
Figure \ref{fig:states}(d) shows a classical ground state phase diagram of the model (\ref{Ham1}).
For negative $J_{\perp}^a$ the anisotropic exchange has the same effect as the long-range
dipolar interactions. It selects the Palmer-Chalker states [\onlinecite{Palmer00}], represented by
the $m_{xy}$ state in Fig.~\ref{fig:states}(a). Their classical energy is
$E_{T_2} = -S^2 (J_{\perp} - \frac{1}{2} J_{\perp}^a)$. For $J_{\perp}^a > 0$ the ground state
belongs to a two component $E$ representation with the energy
$E_{E} = -S^2(J_{\perp} + \frac{1}{2} J_{\perp}^a)$. Its basis is formed by the noncoplanar state
$m_{3z^2-r^2}$ ($\psi_{2}$) and the coplanar $m_{x^2-y^2}$ ($\psi_{3}$) state.
These are shown in Figs.~\ref{fig:states}(b) and \ref{fig:states}(c),  respectively.
The value $J_{\perp}^a=0$ is a highly degenerate point, where many states with different
ordering wavevectors have the same classical energy [\onlinecite{Champion04}].

Focusing on $J_{\perp}^a>0$, we specify the local $\mathbf{\hat{x}}_i$ and
$\mathbf{\hat{y}}_i$ axes on each site along the two $E$-states, see Appendix~\ref{sec:AppSGeom},
and parameterize the whole manifold of degenerate classical ground states with an angle $\varphi$
\begin{equation}
\mathbf{S}_{i} = \mathbf{\hat{x}}_{i}\cos\varphi + \mathbf{\hat{y}}_{i}\sin\varphi.
\label{phi}
\end{equation}
Values $\varphi = \pi k / 3$ and $\varphi = \pi \left( k + \frac{1}{2}\right) / 3$
correspond to different $m_{3z^2-r^2}$ and $m_{x^2-y^2}$ states, respectively.

According to the group theory $m_{3z^{2} - r^2}$ and $m_{x^2-y^2}$ states
remain strictly degenerate for a general case of the bilinear spin Hamiltonian
involving further anisotropic terms or couplings to distant neighbors.
An effective biquadratic exchange $({\bf S}_i \cdot {\bf S}_j)^2$,
does not lift this degeneracy either. The degeneracy may be lifted only by
interactions of the sixth order in spin components, which are usually
small in real materials. Hence, the spin model (\ref{Ham1}) provides an interesting
example of the order from disorder selection. For $0<J_{\perp}^{a}/J_{\perp}<4$,
thermal and quantum fluctuations favor the noncoplanar ground states of the type
$m_{3z^2 - r^2}$, including the point $J_{\perp}^{a}/J_{\perp}\sim 1.5$ corresponding to
$\rm{Er_2Ti_2O_7}$ [\onlinecite{Zhitomirsky12}, \onlinecite{Savary12}, \onlinecite{Zhitomirsky14}].
For $J_{\perp}^{a}/J_{\perp}>4$, the selection takes a different route and
fluctuations stabilize the $m_{x^2-y^2}$ states [\onlinecite{Wong13}].
The corresponding transition at $J_{\perp}^a = 4J_{\perp}$ is indicated
by a dashed line in Fig.~\ref{fig:states}(d).
In the next section we show that quantum and thermal corrections to the classical energy
generate $E_{\rm eff} \sim (J_{\perp}^{a}- 4J_{\perp}) \cos 6\varphi$
explaining the above transition.

\section{Real-space perturbation theory}
\label{sec:RSP}

The aim of this section is to present a simple analytic derivation of the previously
obtained results [\onlinecite{Zhitomirsky12}, \onlinecite{Savary12}, \onlinecite{Champion04},  \onlinecite{Wong13}]
on degeneracy lifting by thermal and quantum fluctuations in the
anisotropic $XY$ pyrochlore. Instead of calculating excitation spectra around a few
selected states, we use the real-space perturbation
theory  [\onlinecite{Maryasin13,Canals04,Heinila93}, \onlinecite{Long89}, \onlinecite{Bergman07}],
which avoids numerical diagonalization and integrations procedures
and treats all possible ground-state spin configurations on
equal footing. Basically, the real-space expansion is a perturbative treatment of
transverse spin fluctuations neglected in the mean-field approximation and, as such, is
a variant of the $1/z$ expansion with $z$ being a number of nearest neighbors (see
Sec.~\ref{SubSQObD}).

\subsection{General formalism}

The real-space perturbation expansion
starts with (i) rewriting the Hamiltonian in
the local frame around an arbitrary ground-state spin configuration and (ii)
separating all terms, which depend on deviation of only one spin. This on-site part
is subsequently regarded as a noninteracting Hamiltonian $\hat{\cal H}_0$ with trivially
calculated excited states. All other terms describe interactions of spin fluctuations
on adjacent sites and are treated as a perturbation $\hat{V}$.
Standard thermodynamic or quantum perturbation theories are used to calculate
the effect of $\hat{V}$. The obtained correction terms generate effective spin-spin
interactions beyond the original spin Hamiltonian and produce the order by disorder
effect. For both quantum and thermal fluctuations,
the second-order expansion generates
effective biquadratic exchange terms  [\onlinecite{Heinila93,Canals04,Maryasin13}].
Here, we need to go to the next third order to obtain effective degeneracy-lifting
interactions in the case of $\rm Er_2Ti_2O_7$.

To proceed with calculations for the aniso\-tropic $XY$ pyrochlore (\ref{Ham1})
we  shall use an alternative form of the spin Hamiltonian
\begin{eqnarray}
\hat{\cal H} & = & \sum_{\langle ij\rangle} \bigl[  -J_{\pm}(S_{i}^{+}S_{j}^{-}
+ S_{i}^{-}S_{j}^{+})
\nonumber \\
 && \phantom{\sum_{\langle i, j\rangle} \bigl[}
 + J_{\pm\pm} (\textnormal{e}^{i\gamma_{ij}}S_{i}^{+}S_{j}^{+}
 + \textnormal{e}^{-i\gamma_{ij}}S_{i}^{-}S_{j}^{-}) \bigr],
\label{Ham2}
\end{eqnarray}
where spin components are assigned for a specific choice of coordinate axes
in the local $xy$ planes, see Appendix~\ref{sec:AppSGeom} for definition
of axes, bond dependent phases $\gamma_{ij}$ and further details.
In particular, the new exchange parameters are related to the original
constants via
\begin{equation}
J_{\pm} = \frac{1}{12}\left(2J_{\perp} + J_{\perp}^{a} \right), \qquad
J_{\pm\pm}=\frac{1}{12}\left(4J_{\perp} - J_{\perp}^{a} \right).
\end{equation}
These new interaction parameters coincide with those used by Savary
{\it et al}. [\onlinecite{Savary12}],
though our spin Hamiltonian is written somewhat differently.

Next we transform to the sublattice basis such that the local $z$-axis becomes parallel
to the spin direction (\ref{phi}) and the local $x$-axis lies in the respective easy plane.
Spin components in the new coordinate frame are denoted by $\textsf{S}^{\alpha}$.
Then, the spin Hamiltonian takes the form
\begin{equation}
\hat{\cal H}= \sum_{\langle ij\rangle}
\bigl[ h_{ij} \textsf{S}_{i}^{z}\textsf{S}_{j}^{z} - M_{ij} \textsf{S}_{i}^{x}\textsf{S}_{j}^{x}
- K_{ij}(\textsf{S}_{i}^{x}\textsf{S}_{j}^{z} + \textsf{S}_{i}^{z}\textsf{S}_{j}^{x})\bigr],
\label{Ham3}
\end{equation}
where $h_{ij}$, $M_{ij}$, and $K_{ij}$ are bond-dependent constants
\begin{eqnarray}
&&h_{ij} = -2J_{\pm} + 2J_{\pm\pm}\cos (2\varphi + \gamma_{ij}),  \nonumber \\
&&K_{ij} = 2J_{\pm\pm} \sin (2\varphi + \gamma_{ij}), \label{hKM} \\
&&M_{ij} = 2J_{\pm} + 2J_{\pm\pm}\cos (2\varphi + \gamma_{ij}).  \nonumber
\end{eqnarray}
They explicitly depend on angle $\varphi$, which parameterizes the classical ground states.
Finally, we extract the on-site part and rewrite (\ref{Ham3}) as
$\hat{\cal H} = \hat{\cal H}_0 + \hat{V}_1 + \hat{V}_2 + \hat{V}_3$, where
\begin{eqnarray}
&&
\hat{\cal H}_0 = h \sum_i (S-\textsf{S}_{i}^{z}),\
\hat{V}_1 = -\!\sum_{\langle ij\rangle} K_{ij} (\textsf{S}_{i}^{x}\textsf{S}_{j}^{z}
+  \textsf{S}_{i}^{z}\textsf{S}_{j}^{x}),
\nonumber  \\
&&
\hat{V}_2 = -\!\sum_{\langle ij\rangle} M_{ij} \textsf{S}_{i}^{x}\textsf{S}_{j}^{x}, \
\hat{V}_3 = \sum_{\langle i j\rangle} h_{ij}(S\!-\textsf{S}^z_i)(S\!-\textsf{S}^z_j).
\label{Ham4}
\end{eqnarray}
The constant $h=\sum_j h_{ij}=12J_{\pm}S$ is an amplitude of a local magnetic field,
which is the same on every site. In the above expression we also omitted a constant
term corresponding to
the classical energy. In the two following subsections we calculate
the relevant energy corrections generated by thermal and quantum fluctuations.

\subsection{Thermal Order by disorder}

Here we consider a  model of purely classical spins of unit length $|{\bf S}_i|=1$.
At low temperatures, spins fluctuate about their equilibrium directions by small
$\textsf{S}^x$ and $\textsf{S}^y$ corresponding to deviations
within the local easy plane and out of it, respectively. The local fluctuations
are governed by
\begin{eqnarray}
\hat{\cal H}_0 = \frac{h}{2}\sum_i (\textsf{S}_{i}^{x}{}^2 + \textsf{S}_{i}^{y}{}^2)\, .
\label{H01}
\end{eqnarray}
The linear in $\textsf{S}^x$ terms included in $\hat{V}_1$ vanish for the lowest-energy
state. Thus, both   $\hat{V}_1$ and $\hat{V}_3$ describe nonlinear effects and
produce higher-order contributions in $T$, which will be neglected in the following.

We now proceed with the classical thermodynamic perturbation theory
to determine the free-energy correction generated by  $\hat{V}_2$.
The calculation is rather straightforward [\onlinecite{Canals04}]
and we present only the final result. The second-order contribution
$\Delta F^{(2)} = -\langle \hat{V}^2\rangle/2T$
is the same for all classical ground states (\ref{phi}).
The leading state-dependent correction appears in the third order:
\begin{eqnarray}
\Delta F^{(3)} = \frac{\langle \hat{V}_2^3\rangle_0}{3T^{2}}
= -\frac{2T}{(12J_{\pm})^{3}}\sum_{\triangle} M_{ij}M_{jk}M_{ki},
\label{dF2}
\end{eqnarray}
where $\langle\ldots\rangle_0$ denotes thermodynamic averaging with respect to
$\hat{\cal H}_0$.
Summation in (\ref{dF2}) is performed over all triangular plaquettes of a pyrochlore
lattice and $i,j,k \in \triangle$.  Substituting $M_{ij}$ from (\ref{hKM}) we obtain
\begin{eqnarray}
\Delta F^{(3)} \simeq - \frac{TJ_{\pm\pm}^{3}N}{216J_{\pm}^{3}}\cos 6\varphi \,.
\label{-cos6f}
\end{eqnarray}
where $N$ is the number of sites. Here and everywhere below $\simeq$ sign means that 
ground state independent constant term has been omitted.
The correction $\Delta F^{(3)}$ is linear in $T$ reflecting
the fact that it is produced by the harmonic fluctuations. It also
has the six-fold symmetry in agreement with the $Z_6$ symmetry breaking in the
$m_{3z^2-r^2}$ magnetic structure [\onlinecite{Zhitomirsky14}]. The respective term
changes sign with $J_{\pm\pm}$, i.e., for $J_{\perp}^a/J_{\perp} =4$, in total agreement
with the phase diagram sketched in Fig.~\ref{fig:states}(d) and with the previous
findings [\onlinecite{Wong13}].
For the ratio of parameters $J_{\perp}^a/J_{\perp} \sim 1.5$ appropriate for
$\rm{Er_{2}Ti_{2}O_{3}}$, $J_{\pm\pm}$ is positive and thermal
fluctuations select $\varphi = 0,\pi/3,...$  corresponding to
the  noncoplanar $m_{3z^{2} - r^2}$ spin configuration.

\subsection{Quantum order by disorder}
\label{SubSQObD}

We now set $T=0$ and use the Rayleigh-Schr\"odinger perturbation theory
to calculate quantum corrections to the classical ground-state energy.
For that we treat $\textsf{S}^{\alpha}$ as spin operators obeying the standard commutation
relations. Again we focus on the effect of $\hat{V}_2$, which is more conveniently
written in terms of spin raising and lowering operators
\begin{equation}
\hat{V}_2 = - \frac{1}{4} \sum_{\langle ij\rangle} M_{ij} (\textsf{S}^+_i+\textsf{S}^-_i)
(\textsf{S}^+_j+\textsf{S}^-_j)\,.
\end{equation}

The ground state $|0\rangle$ of the noninteracting Hamiltonian $\hat{\cal H}_0$
coincides with a chosen classical ground state and corresponds to a `fully-saturated' state
in the rotated basis: $\textsf{S}^+_i |0\rangle= 0$. The latter property yields
$\langle 0|\hat{V}|0\rangle = 0$ and determines that every term in the perturbation
series starts and ends with creation and annihilation of a pair of spin flips.
For instance, the third-order correction is given  by
\begin{equation}
\Delta E^{(3)} = \sum_{n,m} \frac{\langle 0|\hat{V}|n \rangle \langle n | \hat{V} | m \rangle
\langle m | \hat{V} | 0 \rangle}{(E_0 - E_n)(E_0 - E_m)},
\label{dE3}
\end{equation}
where $|n\rangle$ and $|m\rangle$ are excited states of $\hat{\cal H}_0$ with
$E_{n,m} = E_0 + 2h$. Since $h = O(z)$ and $\hat{V}=O(1)$, each extra order of the real-space
expansion contributes a factor $1/z$ to the corresponding energy correction.

Detailed analysis of all second- and third-order terms in the real space perturbation expansion
is presented in Appendix \ref{AppSQ}.  In particular, $\Delta E^{(2)}$ yields an energy shift which is independent
of $\varphi$.  Selection between different ground states is determined by the third-order excitation
process described by the diagram
\begin{equation}
|000\rangle\xrightarrow{\textsf{S}^-_i\textsf{S}^-_j}|110\rangle\xrightarrow{\textsf{S}^+_j\textsf{S}^-_k} |101\rangle
\xrightarrow{\textsf{S}^+_k\textsf{S}^+_i} |000\rangle
\end{equation}
with three sites $i,j,k$ belonging to the same triangular plaquette. The corresponding energy correction
is given by a plaquette sum
\begin{eqnarray}
\Delta E^{(3)} = - 6\sum_\triangle \frac{S^3}{8} \frac{M_{ij}M_{jk}M_{ki}}{(24J_{\pm}S)^2} \,.
\label{Q-cos6f1}
\end{eqnarray}
Performing lattice summation and dropping an unimportant constant we obtain
\begin{eqnarray}
\Delta E^{(3)} \simeq - \frac{J_{\pm\pm}^3NS}{192J_{\pm}^2}\cos 6\varphi\,.
\label{Q-cos6f}
\end{eqnarray}
The  quantum correction scales as $\Delta E^{(3)} = O(JS)$ and, thus, represents a harmonic
spin-wave contribution. The full harmonic spin-wave calculation is, of course, not restricted
to triangular plaquettes and includes graphs of all possible lengths [\onlinecite{Zhitomirsky12}].
However, for small $J_{\pm\pm}/J_\pm$ or $J_\perp^a\agt 1$ its angular dependence as well as
the corresponding prefactor are very closely reproduced by (\ref{Q-cos6f}).

The third-order real-space correction contains also contribution $O(J)$, which goes
beyond the harmonic spin-wave theory. It exhibits the same functional form
as Eq.~(\ref{Q-cos6f}) but has the opposite sign and, therefore, partially compensates
the energy difference between $m_{3z^2-r^2}$ and $m_{x^2-y^2}$ states.
Overall, for $S=1/2$ the amplitude of the sixfold harmonics (\ref{Q-cos6f}) is reduced
by 40\%\ due to interaction effects, see Appendix \ref{AppSQ1/2} for further details.

\section{Order by structural disorder}
\label{sec:structural}

Structural disorder modifies locally exchange interactions and destroys
perfect magnetic frustration at the microscopic level.
As a result, magnetic moments tilt from the equilibrium bulk structure
producing spin textures [\onlinecite{Hoglund07,Eggert07,Wollny11}]
and net  uncompensated moments [\onlinecite{Wollny11,Sen11,Wollny12}].
The idea of uncompensated moments and related local magnetic fields was also used
by Henley  in his explanation of vacancy-induced degeneracy lifting
in the  $J_1$--$J_2$ square-lattice antiferromagnet  [\onlinecite{Henley89}].
Though simple and quite appealing, this approach cannot be applied to
a general problem of `order by structural disorder' in noncollinear
frustrated magnets. Indeed, local fields from vacancies on different magnetic
sublattices average to zero in a macroscopic sample producing no selection.

Building on the previous works [\onlinecite{Slonczewski91}, \onlinecite{Fyodorov91}], we have recently shown that
bond and site disorder generate positive biquadratic exchange [\onlinecite{Maryasin13}].  Such an
effective interaction is obtained by integrating out static fluctuations in a spin texture
and due to its sign favors the {\it least} collinear spin configurations in degenerate
frustrated magnets. Examples include an orthogonal state in the $J_1$--$J_2$ square-lattice
antiferromagnet [\onlinecite{Henley89}, \onlinecite{Weber12}] and a conical state in the Heisenberg triangular
antiferromagnet in an external magnetic field [\onlinecite{Maryasin13}]. Here we extend our treatment of
quenched disorder to the anisotropic $XY$ pyrochlore (\ref{Ham1}). This requires calculation of
the effective Hamiltonian beyond the leading biquadratic contribution.
Also, note that the effect of structural disorder on the equilibrium magnetic structure
is essentially classical. Therefore, we assume throughout this section
that spins are three-component classical vectors with $|{\bf S}_i|=1$.

\subsection{Nonmagnetic impurities}
\label{SubSObSD}

A single vacancy induces a strong local perturbation
of the magnetic structure in noncollinear antiferromagnets [\onlinecite{Wollny11}].
To obtain qualitative insights  within analytic treatment of the impurity problem
 we use a toy model
of weak site disorder [\onlinecite{Fyodorov91}].
Specifically, we let some fraction $n_{\rm imp}$
of classical spins to be shorter by a small amount $\epsilon \ll 1$.
These impurities are distributed randomly over the lattice and we assign a parameter $p_{i}=1$
to every impurity spin and $p_{i}=0$ otherwise: $\sum_i p_i= N_{\rm imp}$.
In the spin Hamiltonian impurities are included by substitution
$\mathbf{S}_{i} \rightarrow \mathbf{S}_{i}(1-\epsilon p_{i})$
and in the leading order in $\epsilon$ we  have for pairwise spin-spin interactions:
\begin{equation}
S_{i}^\alpha S_{j}^\beta \approx S_{i}^\alpha S_{j}^\beta\bigl[1-\epsilon(p_{i} + p_{j}) \bigr].
\end{equation}

We perform the same decomposition of the spin Hamiltonian as described in Sec.~\ref{sec:RSP}.
The main difference with the preceding section is that linear in spin deviations part
 of $\hat{V}_1$ does not vanish:
\begin{equation}
\hat{V}_1^{\prime} = \sum_{\langle i j\rangle} K_{ij} \epsilon \left( p_{i} + p_{j} \right)
(\textsf{S}_{i}^{x} + \textsf{S}_{j}^{x})\,,
\label{V1imp}
\end{equation}
describing the fact that adjacent to impurity spins tilt from their equilibrium orientations
in the bulk.
Minimization of the quadratic form $\hat{\cal H}_0 + \hat{V}_1$ over $\textsf{S}_{i}^{x}$ yields
\begin{eqnarray}
\textsf{S}_{i}^{x} =  \frac{\epsilon}{12J_{\pm}} \sum_{j=1}^6 K_{ij} p_{j}\,,
\label{MinimumC}
\end{eqnarray}
where the sum runs over six nearest neighbors of the site $i$.
Here, we neglected deviations of spins beyond the first-neighbor shell around an impurity
in the spirit of the real-space perturbation expansion.
Substitution of the new minimum condition into $\hat{\cal H}$ produces
an energy correction. The leading term is obtained from
$\hat{\cal H}_0 + \hat{V}_1$ and gives an effective biquadractic exchange [\onlinecite{Maryasin13}].
As before, this energy correction is independent of angle $\varphi$. Going to the next
order we substitute (\ref{MinimumC}) into
$\hat{V}_2$
\begin{equation}
\Delta E_2 = - \frac{\epsilon^{2}}{(12J_{\pm})^{2}}\sum_{\langle ij\rangle} M_{ij} \!
\sum_{l,m=1}^{6}\! K_{il}K_{jm}p_{l}p_{m}\,.
\label{real_vac_minfluct}
\end{equation}
Keeping only terms that are linear in $n_{\rm{imp}}$ we can rewrite Eq.~(\ref{real_vac_minfluct})
as
\begin{equation}
\Delta E_2 =- \frac{\epsilon^{2} n_{\rm{imp}}}{(12J_{\pm})^{2}}\sum_{\langle ij\rangle}
\sum_{l=1}^{2} M_{ij}K_{il}K_{jl}\,,
\end{equation}
where the last summation is over two sites sharing the same tetrahedron with $i$ and $j$.
Finally, substituting expressions for bond-dependent parameters $M_{ij}$ and $K_{ij}$
from (\ref{hKM}) we obtain
\begin{equation}
\Delta E_2 \simeq \frac{J_{\pm\pm}^{3}\epsilon^{2}N_{\rm{imp}}}{12J_{\pm}^{2}} \cos6\varphi \,.
\label{cos6f}
\end{equation}
This energy correction has same symmetry, but the opposite sign compared to
Eqs.~(\ref{-cos6f}) and (\ref{Q-cos6f}). Hence, for $J_{\pm\pm} > 0$ the on-site disorder favors
magnetic configurations (\ref{phi}) with $\varphi = \pi(1+2n)/6$. These correspond to six coplanar
$m_{x^2-y^2}$ states.

\subsection{Bond disorder}

Another type of randomness in magnetic solids is bond disorder. In pyrochlore materials
it may appear as a result of doping on the nonmagnetic $B$ sites. We model this type of
disorder by small random variations of $J_{\pm}$ and $J_{\pm\pm}$:
\begin{eqnarray}
&&J_{\pm}  \longrightarrow J_{\pm}^{ij}=J_{\pm}(1+\delta_{ij})\,, \nonumber \\
&&J_{\pm\pm}  \longrightarrow J_{\pm\pm}^{ij}=J_{\pm\pm}(1+\delta_{ij}) \,.
\label{disorder}
\end{eqnarray}
The fluctuating part $\delta_{ij}$ is assumed to be uncorrelated between
adjacent bonds and relatively small, $\langle \delta_{ij}^2\rangle = D\ll 1$,
such that it does not change the sign of exchange constants.

The subsequent calculation is completely similar to the previous subsection
up to a substitution $\epsilon (p_i + p_j)\rightarrow \delta_{ij}$. The state-dependent
energy correction has the form
\begin{equation}
\Delta E = - \frac{1}{(12J_{\pm})^{2}}\!\sum_{\langle ij\rangle} M_{ij}K_{ij}^{2}\delta_{ij}^{2}
\simeq \frac{J_{\pm\pm}^{3}DN}{24J_{\pm}^{2}} \cos6\varphi .
\label{cos6f1}
\end{equation}

We conclude this section with the remark that a different state selection produced by
structural disorder has its origin in the local breakdown of frustration. Indeed, the corresponding
energy correction is determined by the linear term $\hat{V}_1$, whereas thermal and
quantum order from disorder stems from the quadratic part $\hat{V}_2$. Technically,
the two terms have different combinations of the relative angles
as demonstrated by Eq.~(\ref{hKM}) for the $XY$ pyrochlore antiferromagnet.
In the Heisenberg case there is a similar change between $\hat{V}_1$
and $\hat{V}_2$ consisting in $\sin\theta_{ij}$ and $\cos\theta_{ij}$ prefactors, respectively,
$\theta_{ij}$ being an angle between two spins [\onlinecite{Maryasin13}].
Thus, we may claim that structural disorder has a {\it qualitatively} different effect
on the ground-state selection in a frustrated magnet compared to thermal/quantum fluctuations.
In the purely classical picture the structural disorder always wins over thermal fluctuations
at low temperatures. In real frustrated magnets, the structural disorder must compete with
quantum fluctuations for the state selection at $T=0$. There is a critical strength
of disorder or a critical impurity concentration above which the structural
order from disorder effect  prevails. More detailed consideration of these effects is postponed
till Sec.~VI.

\section{Numerical results}
\label{sec:numerics}

In this section we corroborate the analytic results obtained for weak disorder by numerical
investigation of genuine vacancies in the classical anisotropic $XY$ pyrochlore antiferromagnet.
For that we return back to the original spin Hamiltonian (\ref{Ham1}) and set $J_{\perp} = 1$.
Overall, we present two types of numerical data: determination
of the ground-state magnetic structure at zero temperature and
Monte Carlo simulations of finite-temperature properties. In both cases numerical
computations were performed on periodic clusters of $N=4L^3$ classical spins.
Random vacancies were introduced by setting $|{\bf S}_i|=0$ for a fixed number of
sites $N_{\rm{imp}} = n_{\rm{imp}}N$.  For all computations we employed about 100 independent
impurity configurations used to average numerical data and to estimate the error bars.

For $J^a_\perp>0$, magnetic states of the $XY$ pyrochlore antiferromagnet are characterized by two
order parameters:
\begin{equation}
m = \sqrt{m_x^2 + m_y^2}, \ \ \
m_6 = \frac{1}{m^5}\,\mathrm{Re}\{(m_x + im_y)^6\}.
\label{OPs}
\end{equation}
Here, two components
\begin{equation}
m_x = \frac{1}{N}\sum_i \mathbf{S}_i \cdot \mathbf{\hat{x}}_i\,,\quad
m_y = \frac{1}{N}\sum_i \mathbf{S}_i \cdot \mathbf{\hat{y}}_i
\end{equation}
are defined using a specific choice of axes in the local $xy$ planes,
see Eq.~(\ref{phi}).
Basically, $m$ discriminates ordering within the $E$-manifold
from other irreducible representations of the tetrahedral group, whereas
the clock parameter $m_6 = m \cos 6\varphi$ distinguishes between the
different $E$-states
[\onlinecite{Zhitomirsky14}]. The clock order parameter has a positive
value for six noncoplanar states $m_{3z^2 - r^2}$
and becomes negative for coplanar spin configurations $m_{x^2-y^2}$.

\begin{figure}[t]
\centerline{
\includegraphics[width=0.9\columnwidth]{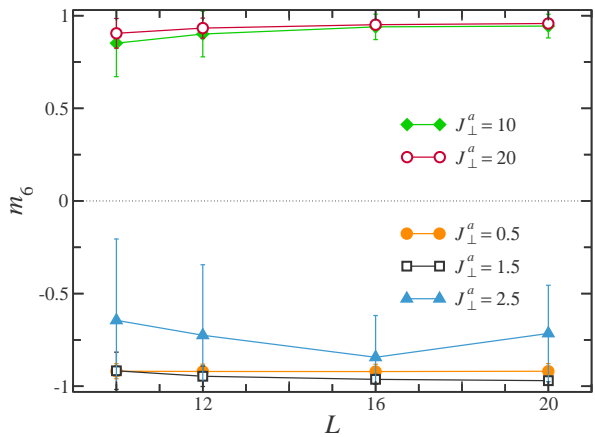}
}
\caption{(Color online) Zero-temperature results for the clock order parameter $m_6$ obtained
for an $XY$ pyrochlore antiferromagnet with $n_{\rm{imp}}=5\%$ of nonmagnetic impurities
for several cluster sizes $L$ and different values of the anisotropic exchange $J^a_\perp$.
Positive and negative values of $m_6$ correspond to $m_{3z^2 - r^2}$ and $m_{x^2-y^2}$ magnetic
states, respectively.
}
\label{fig:GSMin}
\end{figure}

\subsection{Ground state minimization}
\label{sec:GSM}

We begin with minimization of the classical energy (\ref{Ham1}) for a fixed concentration
of static vacancies. We start with a random initial spin
configuration and solve iteratively the classical energy minimum condition
\begin{equation}
\mathbf{S}_{i} =  \mathbf{h}_i/|\mathbf{h}_i|
\end{equation}
with $\mathbf{h}_i$ being the local field on site $i$. After convergence is reached,
the internal energy and the order parameters (\ref{OPs}) are calculated. The process is
repeated for $10^3$ initial random spin configurations and the global minimum is chosen
afterwards. Then, the whole procedure is repeated again for a new configuration of impurities.
The final data are produced by averaging over the lowest energy magnetic structures obtained
for each vacancy set.

Figure \ref{fig:GSMin} shows our results for the clock order parameter $m_6$ in
the $XY$ pyrochlore antiferromagnet with $n_{\rm{imp}}=5\%$ of nonmagnetic impurities.
For each value of the anisotropy parameter $J^a_\perp > 0$ we performed
numerical minimization for several cluster-sizes up to
$L=20$. Negative values of $m_6$ confirm appearance of the coplanar $m_{x^2 - y^2}$ state
induced by impurities for $J^a_\perp<4$. The absolute value of the order parameter grows with
increasing cluster size leaving no doubts about the existence of the true long-range order.
Likewise, for large $J^a_\perp>4$ random impurities stabilize the noncoplanar $m_{3z^2 - r^2}$
magnetic structure characterized by $m_6>0$.

The value $J^a_\perp=4$ ($J_{\pm\pm}=0$) corresponds to isotropic $XY$ spin model in the 
site-dependent local frame. Consequently, two states,
$m_{3z^2 - r^2}$ and $m_{x^2 - y^2}$, remain exactly degenerate for this value of $J^a_\perp$:
neither thermal/quantum fluctuations [\onlinecite{Wong13}] nor impurities (Sec.~\ref{sec:structural})
can lift this degeneracy determined by an emergent $SO(2)$ rotational symmetry of the spin
Hamiltonian. For $J^a_\perp$ close to 4, convergence of the iterative procedure becomes very slow,
see $L=20$ point for $J^a_\perp=2.5$ in Fig.~\ref{fig:GSMin}. One needs to employ a significantly
larger number of initial configurations to approach the true minimum state. This may indicate
the development of some type of glassiness in the system. Similar effect is also
present for very small $J^a_\perp\alt 0.1$ because of additional degeneracy appearing
for $J^a_\perp =0$, see Sec~II. Finally, we studied numerical impurity concentrations
in the range $0.5\% < n_{\rm{imp}} < 7\%$ and obtained the ground state selection independent
of $n_{\rm{imp}}$.

\subsection{Monte Carlo simulations}
\label{sec:MC}

\begin{figure}[t]
\centerline{
\includegraphics[width=0.9\columnwidth]{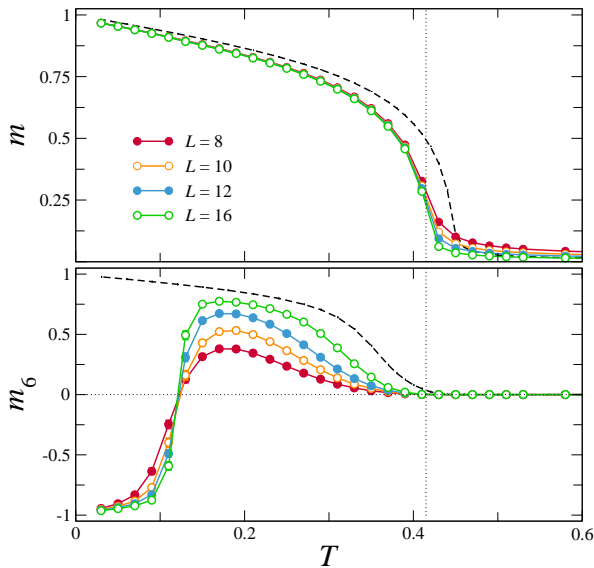}
}
\caption{(Color online)
Monte Carlo results for the antiferromagnetic $m$ (upper panel) and the clock $m_6$
(lower panel) order parameters for the $XY$ pyrochlore  antiferromagnet (\ref{Ham1}) with
$J^{a}_\perp = 0.5$ and $5\%$ of vacancies. Dotted vertical line
indicates the transition temperature. Dashed lines show the behavior of $m$ and $m_6$
for a pure system with $L=16$.}
\label{fig:L}
\end{figure}

Monte Carlo simulations of the classical $XY$ pyrochlore antiferromagnet  were performed
using the Metropolis algorithm alternating five Metropolis steps with five microcanonical
over-relaxation sweeps [\onlinecite{Creutz87}] before every measurement.  In total, $2 \cdot 10^5$
measurements were taken at every temperature and averaging was done over 100 impurity configurations.
We simulated the model (\ref{Ham1}) restricting  variation range of the anisotropy parameter to
$0.3\leq J^a_\perp \leq 2$.

Temperature dependence of the two order parameters $m$ and $m_6$ for $J^a_\perp = 0.5$ and
$n_{\rm{imp}} = 5\%$ is shown in Fig.~\ref{fig:L}. The transition temperature $T_c = 0.415$
was determined from intersection of Binder cumulants $U_L = \langle m^4 \rangle/\langle m^2 \rangle^2$.
It is somewhat reduced compared to the transition into the pure model  $T^0_{c} = 0.4454$ for
the same value of $J^a_\perp$.  The critical behavior of the model (\ref{Ham1}) belongs to
the 3D $XY$ universality class [\onlinecite{Zhitomirsky14}] with the known value of the correlation
length exponent $\nu\approx 0.672$ [\onlinecite{Campostrini06}]. We can now use the Harris criterion
[\onlinecite{Harris74}], which states that the critical behavior for phase transitions with
$d\nu >2$ remains unchanged in the presence of disorder. Since $\nu$ is slightly larger than 2/3,
the critical point in the $XY$ pyrochlore antiferromagnet remains unaffected upon dilution
with nonmagnetic impurities.

Nevertheless, the diluted antiferromagnet exhibits the  peculiar temperature dependence of
the clock order parameter $m_6$, see the lower panel of Fig.~\ref{fig:L}.
Right below $T_c$, $m_6$ is positive, as expected
for the $m_{3z^2-r^2}$ state, and  grows at fixed $T$ with the system size $L$.
Such `inverse' finite-size scaling is attributed to the dangerously irrelevant
role of the six-fold anisotropy at the $XY$ transition in three dimensions and is explained by presence
of an additional length-scale $\xi_6\gg\xi$ [\onlinecite{Zhitomirsky14}].
Upon further cooling, the clock order parameter shows a sharp jump to negative values at $T_1\approx 0.12$.
This jump signifies a phase transition into the $m_{x^2-y^2}$ state stabilized
by impurities. Basically, the temperature dependence of $m_6$ is determined by competition of two terms:
the impurity correction $\Delta E_2$ given by Eq.~(\ref{cos6f}) and the free-energy
correction $\Delta F^{(3)}$ generated by thermal fluctuations (\ref{-cos6f}). They have different sign
and at $T\to 0$ the impurity contribution dominates selecting the $m_{x^2-y^2}$ state.
However, thermal fluctuations grow with temperature and above $T_1$ the effective anisotropy $\Delta F^{(3)}$
wins over $\Delta E_2$ leading to the $m_{3z^2-r^2}$ state right below $T_c$.

\begin{figure}[b]
\centerline{
\includegraphics[width=0.9\columnwidth]{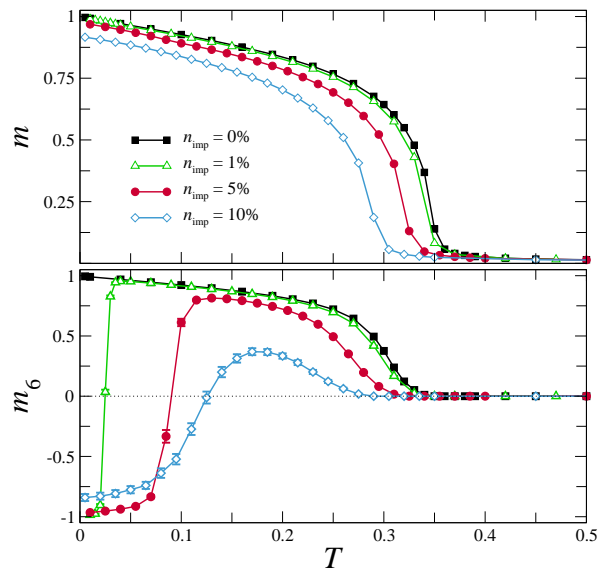} \hspace{2mm}
}
\caption{(Color online)
Temperature dependence of the magnetic order parameters for different impurity
concentrations. Monte Carlo results are for
$J^a_\perp = 0.3$ and $L=16$.}
\label{fig:Nimp}
\end{figure}

The phase transition between $m_{3z^2-r^2}$  and  $m_{x^2-y^2}$ states
is expected to be of the first order on symmetry grounds. (Another possibility is two closely
located second-order transitions with an intermediate low-symmetry phase.)
We collected histograms for the clock order parameter $m_6$ for a few impurity concentrations,
which confirm the first-order nature of the transition. On the other hand, no anomaly is
seen in the specific heat or magnetic susceptibility even for the largest clusters.
A similar behavior was also observed in our previous study of the triangular Heisenberg antiferromagnet
with vacancies [\onlinecite{Maryasin13}].
Thermodynamic signatures of the first-order transition appear to be blurred by disorder.

The observed sequence of ordered phases remains stable under variations of $n_{\rm{imp}}$
and $J^a_\perp$. Figure~\ref{fig:Nimp} shows dependence on vacancy concentration for
$J^a_\perp = 0.3$. We include only Monte Carlo results for the largest clusters with $L=16$.
The first-order transition temperature progressively grows between $T_1=0.025$ for $n_{\rm imp}=1$\%
to $T_1=0.125$ for $n_{\rm imp}=10$\%. The order parameter jump is very sharp for the lowest impurity
concentration but becomes significantly smeared for $n_{\rm imp}=10$\%.  We attribute this effect
to a substantial finite-size scaling at large impurity concentrations. Monte Carlo simulations of
significantly bigger clusters are required for precise determination of the transition point between
the two $E$ states for large density of vacancies.

\begin{figure}[t]
\centerline{
\includegraphics[width=0.9\columnwidth]{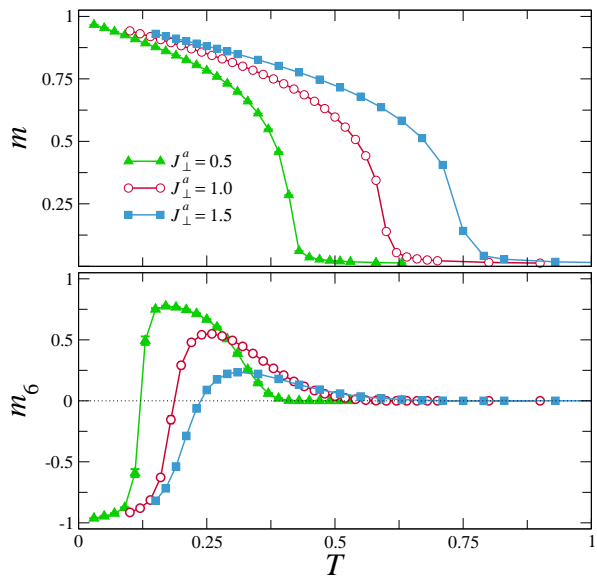}
}
\caption{(Color online)
Temperature dependence of the magnetic order parameters for fixed impurity concentration
$n_{\rm{imp}}=5\%$ and different values of $J^a_\perp$. Monte Carlo results are for
$L=16$ clusters.}
\label{fig:Ja}
\end{figure}

Finally, dependence on $J^a_\perp$ is illustrated in Fig.~\ref{fig:Ja}. As expected, the
$m_{x^2-y^2}$ state is present at low temperatures for all studied values of the anisotropic
exchange including $J^a_\perp=1.5$, which is very close to the experimental estimate for
$\rm Er_2Ti_2O_7$, see Appendix~\ref{sec:AppSGeom}. Somewhat surprisingly,  the thermal selection
of the $m_{3z^2-r^2}$ state in the vicinity of $T_c$ is also remarkably stable under variations of $n_{\rm imp}$
or $J^a_\perp$. This can be considered as a consequence of the Harris criterion, which
asserts irrelevance of quenched disorder for transitions in the 3D $XY$ universality class.

\section{Structural vs quantum disorder}

The Monte Carlo results of the previous section give a general idea about competition
between impurities and thermal fluctuations for the ground-state selection. Beyond the 
classical model, similar competition exists also for the structural disorder and quantum 
effects even at zero temperature. There must be a critical impurity concentration above 
which the quantum selection gives way to the spin configurations stabilized by vacancies.
Since the energy gain produced by impurity substitution is a purely classical effect,
the critical impurity concentration  scales with the spin length as $n_c \sim 1/S$.
This raises a legitimate question about observability of the structural order from disorder
effect in spin-1/2 frustrated magnets, in particular, in diluted $\rm Er_2Ti_2O_7$.

Even an approximate calculation of $n_c$ is a fairly difficult theoretical problem.
In order to treat quantum effects within the framework of the $1/S$ spin-wave expansion,
one starts with setting up the Holstein-Primakoff transformation from spins to bosons in
the local frame around a specific magnetic state. Subsequent calculation of the quantum
energy correction  can be performed directly in the real space without a need to do
the Fourier transformation [\onlinecite{Wessel05,Tcher04}].
Then, in full analogy with Sec.~\ref{sec:GSM}, one can find numerically the harmonic energy
correction for an  impurity-induced nonuniform
spin texture and average that over random impurity configurations.
However, a similar computation for the competing spin configurations
selected by quantum fluctuations at $n_{\rm imp}=0$ immediately fails. Such states
cease to be the classical ground states in the presence of impurities and, hence, have ill-defined
harmonic excitation spectra. The quantum order by disorder selection is manifestly nonlinear
effect in the presence of structural randomness.

Here, we circumvent the difficulty of treating nonlinear quantum effects for an
impurity-induced spin texture in $\rm Er_2Ti_2O_7$, by assuming that concentration of
vacancies is low. Then, the quantum energy correction can be taken as that for the
pure $S=1/2$ pyrochlore antiferromagnet (\ref{Q-cos6f}), whereas the classical energy
gain from impurities is estimated from Eq.~(\ref{cos6f}) by restoring the $S^2$ prefactor
and substituting $\epsilon=1$. Actually, instead of the harmonic result (\ref{Q-cos6f}) we
employ a more accurate expression (\ref{AppQ-cos6f1/2}) with a 40\%\ reduced amplitude
for the sixfold harmonics due to renormalization by interaction effects. In this way we
obtain a reasonably small value of the critical impurity concentration $n_c \approx 7\%$
being only weakly dependent on the ratio of $J_{\pm\pm}/J_\pm$.
For comparison, the magnetization plateau in the Heisenberg triangular antiferromagnet
remains stable up to  $n_c \sim 20\%$ for $S=1/2$ basically meaning that the dilution
effects in this case are observable only for large spins $S\agt 1$ [\onlinecite{Maryasin13}].

Undoubtedly, the above estimate is rather crude and there are good chances that 
the critical impurity concentration for $\rm Er_2Ti_2O_7$ is even smaller than 7\%.
The approximation adopted for derivation of (\ref{cos6f}) treats only tilting
of nearest-neighbor spins around a vacancy. Inclusion of full-range spin relaxation
in an impurity-induced magnetic texture should further increase the corresponding energy gain
and, hence, reduce the critical value of $n_{\rm imp}$.
Thus, we may conclude that quantum effects become subdominant in the anisotropic $XY$ pyrochlore
already for small
dilution and there is a good prospective for an experimental observation of the impurity
induced $m_{x^2-y^2}$ state in $\rm Er_2Ti_2O_7$.

\section{Conclusions}
\label{sec:disc}

To summarize, we have studied the effect of nonmagnetic dilution and weak bond disorder
for the anisotropic $XY$ pyrochlore antiferromagnet.  The degeneracy lifting produced
by the two types of disorder is opposite to the effect of thermal and quantum fluctuations.
Specifically, in the parameter range $J_\perp^a/J_\perp<4$ relevant for $\rm Er_2Ti_2O_7$,
the structural disorder stabilizes at zero temperature
the coplanar $m_{x^2-y^2}$ magnetic structure. At finite temperatures, thermal fluctuations
induce the reentrant first-order transition into the $m_{3z^2-r^2}$ state.
Our results further confirm the striking dissimilarity  between the order from disorder effects
generated by fluctuations, thermal or quantum, and by frozen disorder in the spin Hamiltonian
parameters.
In a broader prospective, the different ground-state selection is produced by different
coupling to transverse spin fluctuations and, therefore, should persists for various generalizations
of the spin Hamiltonian including frustrated spin-orbital models [\onlinecite{Khaliullin05}].
Finally, let us remark that completely unambiguous identification of the quantum order from disorder
effect always meets a problem of  distinguishing  it from weak extra interactions like, for example,
spin-lattice coupling in the case of the magnetization plateaus [\onlinecite{Penc04}].
On the other hand, controlled doping of nonmagnetic impurities into a frustrated magnet
may provide a clear experimental evidence of the structural order from disorder phenomenon.

\acknowledgments
We are grateful to R. Moessner and M. Vojta for fruitful discussions.

\appendix
\section{Spin Hamiltonian}
\label{sec:AppSGeom}

In cubic pyrochlore materials, magnetic ions form a network of corner-sharing tetrahedra usually
called a pyrochlore lattice. The unit cell contains four magnetic sites.
Their positions in units of the cubic lattice parameter $a$ and the directions of local
$\langle 111\rangle$ axes are given by
\begin{align}
&\mathbf{r}_1 = (0,0,0)\,, \quad && \hat{\bf z}_1 = \tfrac{1}{\sqrt{3}}(1,1,1)\,, \nonumber \\
&\mathbf{r}_2 = (\tfrac{1}{4},\tfrac{1}{4},0)\,, \quad && \hat{\bf z}_2 = \tfrac{1}{\sqrt{3}}(-1,-1,1)\,,\nonumber \\
&\mathbf{r}_3 = (0,\tfrac{1}{4},\tfrac{1}{4})\,, \quad && \hat{\bf z}_3 = \tfrac{1}{\sqrt{3}}(1,-1,-1)\,, \\
&\mathbf{r}_4 = (\tfrac{1}{4},0,\tfrac{1}{4})\,, \quad && \hat{\bf z}_4 = \tfrac{1}{\sqrt{3}}(-1,1,-1)\,.\nonumber
\label{AppZaxis}
\end{align}
The primitive lattice vectors are chosen as $\mathbf{a}_1 = (\frac{1}{2},\frac{1}{2},0)$,
$\mathbf{a}_2 = (0,\frac{1}{2},\frac{1}{2})$, and  $\mathbf{a}_3 = (\frac{1}{2},0,\frac{1}{2})$.

The most general form of the anisotropic exchange Hamiltonian
for pseudo-spin-1/2 operators representing erbium magnetic moments can be
written as [\onlinecite{Zhitomirsky12}, \onlinecite{Zhitomirsky14}]
\begin{eqnarray}
\hat{\cal H} & = & \sum_{\langle ij\rangle}  J_{zz} S_{i}^{z}S_{j}^{z} +
J_{\perp} {\bf S}^{\perp}_{i}\cdot{\bf S}^{\perp}_{j} + J_{\perp}^{a}
({\bf S}^{\perp}_{i}\cdot \hat{\bf r}_{ij})({\bf S}^{\perp}_{j}\cdot \hat{\bf r}_{ij}) \nonumber \\
 && \phantom{\sum_{\langle ij\rangle}}  + J_{z\perp}
\bigl[S_{j}^{z}({\bf S}^{\perp}_{i}\cdot \hat{\bf r}_{ij}) + S_{i}^{z}({\bf S}^{\perp}_{j}\cdot \hat{\bf r}_{ji})\bigr].
\label{AppHam0}
\end{eqnarray}
Here $\hat{\bf r}_{ij} = ({\bf r}_{i} - {\bf r}_{j}) / |{\bf r}_{i} - {\bf r}_{j}|$ is a unit vector in the bond direction.
Spin operators $\mathbf{S}_{i}$ are taken in the local coordinate frame with $S_{i}^{z}$ and $\mathbf{S}^{\perp}$
being projections on the local trigonal axis and on the orthogonal $xy$ plane, respectively.
Being independent of the choice of $x$ and $y$ axes, this form of the Hamiltonian
is convenient for calculation of classical energies and for Monte Carlo simulations.

To describe the classical ground states of the $XY$ pyrochlore antiferromagnet we choose directions of
$\hat{\bf x}_i$ and $\hat{\bf y}_i$ axes such that they coincide with sublattice direction for
the $m_{3z^{2} - r^{2}}$ and the $m_{x^2-y^2}$ state, respectively:
\begin{align}
&\mathbf{\hat{x}}_1 = \tfrac{1}{\sqrt{6}}(1,1,-2)\,, \quad && \mathbf{\hat{y}}_1 = \tfrac{1}{\sqrt{2}}(-1,1,0)\,,   \nonumber \\
&\mathbf{\hat{x}}_2 = \tfrac{1}{\sqrt{6}}(-1,-1,-2)\,, \quad && \mathbf{\hat{y}}_2 = \tfrac{1}{\sqrt{2}}(1,-1,0)\,, \nonumber  \\
&\mathbf{\hat{x}}_3 = \tfrac{1}{\sqrt{6}}(1,-1,2)\,, \quad && \mathbf{\hat{y}}_3 = \tfrac{1}{\sqrt{2}}(-1,-1,0)\,,  \\
&\mathbf{\hat{x}}_4 = \tfrac{1}{\sqrt{6}}(-1,1,2)\,, \quad && \mathbf{\hat{y}}_4 = \tfrac{1}{\sqrt{2}}(1,1,0)\,.\nonumber
\label{AppAxes}
\end{align}

The spin Hamiltonian used in Sec.~\ref{sec:RSP} is derived from an alternative form of the spin Hamiltonian (\ref{AppHam0}):
\begin{eqnarray}
\hat{\cal H}  & = & \sum_{\langle ij\rangle}\Bigl\{ J_{zz} S_i^zS_j^z -
J_\pm (S^+_i S^-_j + S_i^-S_j^+) \nonumber \\
&& \phantom{\sum_{\langle ij\rangle}} + J_{\pm\pm} (\textnormal{e}^{i\gamma_{ij}}S^+_i S^+_j + \textnormal{e}^{-i\gamma_{ij}}S_i^-S_j^-) \\
&& \phantom{\sum_{\langle ij\rangle}} - J_{z\pm} \bigl[S_j^z(\textnormal{e}^{-i\gamma_{ij}}S^+_i +\textnormal{e}^{i\gamma_{ij}}S_i^-)
+ i \leftrightarrow j\bigr] \Bigr\}\,, \nonumber
\label{AppHsav}
\end{eqnarray}
were phases $\gamma_{ij}$ explicitly depend on the choice of basis in the $xy$ planes.
This form of the spin Hamiltonian was previously employed in a number of works [\onlinecite{Savary12}, \onlinecite{Onoda10}, \onlinecite{Ross11}]
with a minor redefinition of complex factors. Instead of using $\gamma_{ij}
\equiv \textnormal{e}^{i\gamma_{ij}}$ and $\zeta_{ij} \equiv -\textnormal{e}^{-i\gamma_{ij}}$,
we explicitly extract phases, which greatly simplifies our subsequent expressions.
For the above choice of axes we have
\begin{equation}
\gamma_{12} = \gamma_{34} = 0\,, \ \  \gamma_{13} = \gamma_{24} = -\gamma_{14} = -\gamma_{23} =\frac{2\pi}{3}\,.
\label{Appphases}
\end{equation}
Comparing two forms of the spin Hamiltonian  we obtain the following relation between
two sets of exchange parameters:
\begin{eqnarray}
&& J_\perp = 2(J_\pm + J_{\pm\pm})\,, \quad
J^a_\perp = 8J_\pm - 4J_{\pm\pm}\,, \nonumber \\
&&
J_{z\perp} = -2\sqrt{3}J_{z\pm}\,.
\label{Jconv}
\end{eqnarray}

Neutron measurements of magnetic excitations in $\rm Er_2Ti_2O_7$
in a high magnetic field [\onlinecite{Savary12}], yield the following estimate for the exchange
parameters in $10^{-2}$~meV:
\begin{eqnarray}
&& J_\pm = 6.5\pm0.75\,, \quad \ \  J_{\pm\pm} = 4.2\pm0.5\,, \nonumber \\
&& J_{zz} = -2.5\pm1.8\,, \quad J_{z\pm} = -0.88\pm1.5~\,.
\end{eqnarray}
Applying (\ref{Jconv}) we obtain
\begin{eqnarray}
&& J_\perp = 0.21(2)~\textrm{meV}\,, \qquad \ \ J^a_\perp = 0.35(5)~\textrm{meV}\,, \nonumber \\
&& J_{zz} = -0.025(2)~\textrm{meV}\,, \quad J_{z\perp} = 0.03(5)~\textrm{meV}\,.
\end{eqnarray}
The above values confirm the planar character of the interaction
between Er$^{3+}$ moments as well as a significant anisotropy for the in-plane exchange
constants $J_\perp^a/J_\perp \approx 1.7$. Since transverse exchange constants $J_{zz}$ and
$J_{z\perp}$ are an order of magnitude smaller, the physical properties
$\rm Er_2Ti_2O_7$ can be quite accurately captured  by the model with just two exchange
parameters $J_\perp$ and $J^a_\perp$ or $J_\pm$ and $J_{\pm\pm}$ used in the main text.

\section{Quantum order by disorder}
\label{AppSQ}

Here we provide more details on the calculation of the quantum correction (\ref{Q-cos6f}),
which selects between different classical spin configurations. The basic set up of the
perturbation expansion is described in Sec.~III, see Eqs.~(\ref{hKM}) and (\ref{Ham4}).
The ground state $|0\rangle$ of the noninteracting Hamiltonian $\hat{\cal H}_0$
coincides with a classical state and corresponds to a fully-saturated state
in the local basis: $\textsf{S}^+_i |0\rangle= 0$. As a result,
the perturbation series starts with the second-order correction
\begin{equation}
\Delta E^{(2)} = \sum_n \frac{\langle 0|\hat{V}|n \rangle \langle n | \hat{V} | 0 \rangle}
{E_0 - E_n}\,.
\label{dE2}
\end{equation}
Due to the specific form of the perturbation $\hat{V} = V_2 + V_3$, the intermediate excited states
$|n\rangle$ have only two spin flips on neighboring lattice sites
with $E_n = E_0 + 2h$. These processes are represented
by the following diagram:
\begin{eqnarray}
|00\rangle\xrightarrow{\textsf{S}^-_i\textsf{S}^-_j}|11\rangle\xrightarrow{\textsf{S}^+_i\textsf{S}^+_j}|00\rangle\,,
\label{AppQprocess2}
\end{eqnarray}
which describes creation and subsequent annihilation of a pair of spin flips on the same bond $\langle ij\rangle$.
The energy correction from these processes is
\begin{equation}
\Delta E^{(2)}=-\sum_{\langle ij\rangle} \frac{S^2}{4}\frac{M_{ij}^2}{2h}
= -\frac{NSJ_\pm}{8}\Bigl(1+\frac{J_{\pm\pm}^2}{2J_\pm^2}\Bigr).
\label{dE2a}
\end{equation}
Absence of $\varphi$-dependence in the final expression can be easily understood by noticing that
$M_{ij}\propto\cos 2\varphi$ and, consequently,  $\Delta E^{(2)}\propto\cos 4\varphi$.
However, all lower-order $\varphi$-harmonics are prohibited by the $Z_6$ symmetry
with the proper angular-dependent contribution  $\Delta E\sim M_{ij}^3\propto \cos 6\varphi$ arising
only in the third order of the perturbation expansion.

The third-order energy correction is given by the Rayleigh-Schr\"odinger
expression (\ref{dE3}). In this order, there are two distinct excitation processes.
The first one considered in the main text is represented by the triangular plaquette
diagram
\begin{equation}
|000\rangle\xrightarrow{\textsf{S}^-_i\textsf{S}^-_j}|110\rangle\xrightarrow{\textsf{S}^+_j\textsf{S}^-_k} |101\rangle
\xrightarrow{\textsf{S}^+_k\textsf{S}^+_i} |000\rangle\,.
\end{equation}
The full expression of the corresponding energy correction is given by (\ref{Q-cos6f1}) and
 after summation transforms into
\begin{eqnarray}
\Delta E^{(3a)}\! = - \frac{J_{\pm\pm}^3NS}{192J_{\pm}^2}\cos 6\varphi-
\frac{NSJ_\pm}{48}\Bigl(1\! - \frac{3J_{\pm\pm}^2}{4J_\pm^2}\!\Bigr).\qquad
\label{AppQ-cos6f}
\end{eqnarray}

\begin{figure}[t]
\centerline{
\includegraphics[width=0.9\columnwidth]{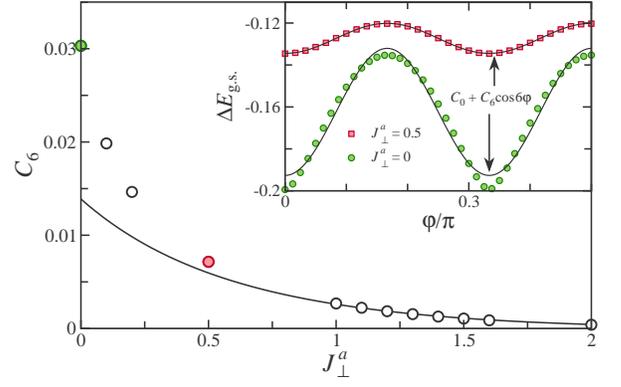}
}
\caption{(Color online)
Inset: fit of the spin-wave energy correction $\Delta E_{\rm g.s.}$ 
(data points) to $\cos 6\varphi$ angular dependence (full lines).
Main panel: comparison of the fitted amplitude $C_6$ of the cosine function
(circles) to the analytic expression (solid line) from the leading term in the harmonic 
expansion (\ref{AppQ-cos6f}).
}
\label{fig:QQ}
\end{figure}

The scaling $\Delta E^{(3a)}\sim JS$ indicates that this correction is one of the terms included
into the harmonic spin-wave theory. Thus, we may directly compare the analytic expression (\ref{AppQ-cos6f})
to the full ground-state energy correction of the harmonic spin-wave theory,
which requires numerical integration of magnon energies over the Brillouin zone.
Results are presented in Fig.~\ref{fig:QQ}.
The inset shows  $\Delta E_{\rm g.s.}$
for $J_\perp^a/J_\perp = 0$ and 0.5 (data points)
from Ref.~\onlinecite{Zhitomirsky12} and fits to
\begin{equation}
\Delta E_{\rm g.s.}= C_0 + C_6 \cos 6\varphi
\label{CA}
\end{equation}
dependence (full lines). The $J_\perp^a/J_\perp = 0$ data exhibit the required
sixfold periodicity but noticeably deviate from a simple cosine function.
On the other hand, for $J_\perp^a/J_\perp = 0.5$ and larger, the spin-wave data are perfectly fitted
by Eq.~(\ref{CA}).  Values of $C_6$ obtained from the fit are compared to the analytic expression
(\ref{AppQ-cos6f}) in the main panel of Fig.~\ref{fig:QQ}. The agreement is very good for
$J_\perp^a/J_\perp \agt 1$ signaling that accuracy of the analytic result is basically
governed by a small ratio
\begin{equation}
\frac{J_{\pm\pm}}{J_\pm} = \frac{4J_\perp-J_\perp^a}{2J_\perp+J_\perp^a}\,.
\end{equation}
We have also compared the
constant $C_0$ to $\varphi$-independent contributions in Eqs.~(\ref{dE2a}) and (\ref{AppQ-cos6f})
and found agreement within 15--25\%, which is a reasonable accuracy for just two first terms of
the $1/z$ expansion.

Another third-order contribution is described by the diagram
\begin{equation}
|00\rangle\xrightarrow{\textsf{S}^-_i\textsf{S}^-_j} |11\rangle \xrightarrow{\textsf{S}^z_i\textsf{S}^z_j}
|11\rangle \xrightarrow{\textsf{S}^+_i\textsf{S}^+_j} | 00 \rangle
\label{AppQprocess3b}
\end{equation}
and corresponds to a single-bond process providing correction to (\ref{dE2a})
due to interaction of two spin flips generated by $\hat{V}_3$. Its explicit expression is
\begin{equation}
\Delta E^{(3b)} = \sum_{\langle ij\rangle} \frac{S^2}{4} \frac{M_{ij}^2h_{ij}}{(24J_{\pm}S)^2}\,.
\end{equation}
This energy correction also has a $\varphi$-dependent part:
\begin{equation}
\Delta E^{(3b)} \simeq \frac{NJ_{\pm\pm}^3}{384J_{\pm}^2}\cos 6 \varphi.
\label{AppQcos6f}
\end{equation}
The coefficient in front of the cosine function is positive providing an opposite
tendency compared to $\Delta E^{(3a)}$. Still,
for $S \geq 1$ and $J_{\pm\pm}>0$ the total third-order correction
robustly selects the $m_{3z^2-r^2}$ states. However, for the case $S=1/2$
relevant to $\rm Er_2Ti_2O_7$, the two angular-dependent terms  cancel
each other and one needs to perform a more careful analysis of the real-space
perturbation terms. We present further details on that in the next Appendix, though the main conclusion
on the state selection by quantum fluctuations remains intact.

\section{Quantum perturbation theory for $S=1/2$}
\label{AppSQ1/2}

In Appendix \ref{AppSQ} we have found that the third-order quantum correction $\Delta E^{(3b)}$
resulting from interaction of two spin flips cancels the harmonic spin-wave contribution
$\Delta E^{(3a)}$ leaving intact the degeneracy between $m_{3z^2-r^2}$ and $m_{x^2-y^2}$
states. To treat more carefully interaction effects we adopt a modified real-space expansion
based on a partial rearrangement of
perturbation terms in Eq.~(\ref{Ham4}). Specifically, $\hat{V}_3$ is now included into a new
unperturbed Hamiltonian
\begin{equation}
\hat{\cal H}_0^{\prime} = h\sum_i \left(S-\textsf{S}^z_i \right)
 + \sum_{\langle ij\rangle}  h_{ij}(S-\textsf{S}^z_i)(S-\textsf{S}^z_j)\,.
\label{AppHq1/2}
\end{equation}
In this way the Ising part of spin-flip interaction is treated exactly. Basically, the new expansion
corresponds to resummation of an infinite subset of terms in the original real-space approach used
in Sec.~III and Appendix~\ref{AppSQ}. A similar trick was also applied in Ref.~\onlinecite{Bergman07}
for the Ising expansion at the fractional magnetization plateaus.

The main difference between the two forms of the real-space expansion is in assignment of
excitation energies in (\ref{dE2}). The lowest-energy excitation, a pair of spin-flips on the same bond,
costs now $E_n - E_0 = 2h + h_{ij}$. We rewrite it as
\begin{equation}
E_n - E_0 = 2J_{\pm}\varepsilon + 2J_{\pm\pm}\cos(2\varphi + \gamma_{ij})\,,
\label{dEij}
\end{equation}
with $\varepsilon = (12S-1)$ and use $1/\varepsilon$ as a small parameter.

The second-order energy correction corresponds to single-bond processes and is expressed by
\begin{equation}
\Delta E^{(2)} = -\frac{S^2}{4} \sum_{\langle ij\rangle} \frac{M_{ij}^2}{2h + h_{ij}}\,.
\end{equation}
Keeping only the lowest-order terms up to $O\left(\varepsilon^{-3}\right)$ and dropping all
unessential constants we obtain
\begin{equation}
\Delta E^{(2)} \simeq \frac{3}{8}\frac{NS^2J_{\pm\pm}^3}{J_{\pm}^2\varepsilon^3}
(\varepsilon - 2) \cos6\varphi \,.
\end{equation}
For large $S$ this expression matches exactly with the corresponding term in $\Delta E^{(3b)}$, see (\ref{AppQcos6f}).

In the third-order, there is only a triangular-plaquette process, which provides the energy shift
\begin{equation}
\Delta E^{(3)} = -6\sum_{\triangle} \frac{S^3}{8}\frac{M_{ij}M_{ki}M_{kj}}{(2h + h_{ij})(2h + h_{ik})} \,.
\end{equation}
We calculate it expanding in powers of $1/\varepsilon$ as
\begin{equation}
\Delta E^{(3)} \simeq - \frac{N}{16}\frac{S^2J_{\pm\pm}^3}{J_{\pm}^2\varepsilon^3} \Bigl(\varepsilon^2 +
\varepsilon + \frac{3}{4}\frac{J_{\pm\pm}^2}{J_{\pm}^2}\Bigr)\cos6\varphi\,.
\end{equation}
The first leading term again matches the angular-dependent part of the
previous expression (\ref{AppQ-cos6f}).

Comparing now $\Delta E^{(2)}$ and $\Delta E^{(3)}$, we do see cancellation of the leading
$1/\varepsilon$ terms for $S=1/2$. Still the coefficient in front of the cosine is negative:
\begin{equation}
\Delta E = - \frac{3}{2000}\frac{NJ_{\pm\pm}^3}{J_{\pm}^2}
\Bigl(1+\frac{1}{16}\frac{J_{\pm\pm}^2}{J_{\pm}^2}\Bigr)\cos 6\varphi\,.
\label{AppQ-cos6f1/2}
\end{equation}
Thus, for $S=1/2$ the quantum order by disorder selection acts in the same way as for large spins.
The main consequence of spin flip interactions is  $\sim 40$\% reduction of the amplitude
of the cosine harmonics as compared to the noninteracting result (\ref{Q-cos6f}) or (\ref{AppQ-cos6f}).


\begin{thebibliography}{99}

\bibitem{Villain80}
J. Villain, R. Bidaux, J.-P. Carton, and R Conte,
J. de Physique \textbf{41}, 1263 (1980).

\bibitem{Shender82}
E. F. Shender, Zh. Eksp. Teor. Fiz. \textbf{83}, 326 (1982)
[Sov. Phys. JETP \textbf{56}, 178 (1982)].

\bibitem{Inami96}
T. Inami, Y. Ajiro, and T. Goto,
J. Phys. Soc. Jpn. \textbf{65}, 2374 (1996).

\bibitem{Smirnov07}
A. I. Smirnov, H. Yashiro, S. Kimura, M. Hagiwara, Y. Narumi, K. Kindo,
A. Kikkawa, K. Katsumata, A. Ya. Shapiro, and L. N. Demianets,
Phys. Rev. B {\bf 75}, 134412 (2007).

\bibitem{Fortune09}
N. A. Fortune, S. T. Hannahs, Y.  Yoshida, T. E. Sherline,
T. Ono, H. Tanaka,  and Y. Takano, Phys. Rev. Lett.  \textbf{102}, 257201 (2009).

\bibitem{Susuki13}
T. Susuki, N. Kurita, T. Tanaka, H. Nojiri, A. Matsuo, K. Kindo, and H. Tanaka,
Phys. Rev. Lett. {\bf 110}, 267201 (2013).

\bibitem{Zhitomirsky12}
M. E. Zhitomirsky, M.V. Gvozdikova, P.C.W. Holdsworth, and R. Moessner,
Phys. Rev. Lett. {\bf 109}, 077204 (2012).

\bibitem{Savary12}
L. Savary, K. A. Ross, B. D. Gaulin, J. P. C. Ruff, and L. Balents,
Phys. Rev. Lett. {\bf 109}, 167201  (2012).

\bibitem{Henley89}
C. L. Henley,
Phys. Rev. Lett. \textbf{62}, 2056 (1989).

\bibitem{Fyodorov91}
Y. V. Fyodorov and E. F. Shender,
J. Phys.: Condens. Matter \textbf{3}, 9123 (1991).

\bibitem{Weber12}
C. Weber and F. Mila,
Phys. Rev. B \textbf{86}, 184432 (2012).

\bibitem{Maryasin13}
V. S. Maryasin and M. E. Zhitomirsky,
Phys. Rev. Lett. {\bf 111}, 247201, (2013).

\bibitem{Heinila93}
M. T. Heinil\"a and A. S. Oja,
Phys. Rev. B {\bf 48}, 7227 (1993).

\bibitem{Canals04}
B. Canals and M. E. Zhitomirsky,
J. Phys.: Condens. Matter {\bf 16}, S759 (2004).

\bibitem{Slonczewski91}
J. C. Slonczewski,
Phys. Rev. Lett. {\bf 67}, 3172 (1991).

\bibitem{Champion03}
J. D. M. Champion, M. J. Harris, P. C. W. Holdsworth, A. S. Wills, G. Balakrishnan,
S. T. Bramwell, E. \v{C}i\v{z}m\'{a}r, T. Fennell, J. S. Gardner, J. Lago, D. F. McMorrow,
M. Orend\'{a}\v{c}, A. Orend\'{a}\v{c}ov\'{a}, D. McK. Paul, R. I. Smith,
M. T. F. Telling, and A. Wildes, Phys. Rev. B \textbf{68}, 020401(R) (2003).

\bibitem{Poole07}
A. Poole, A. S. Wills, and  E. Leli\`evre-Berna,
J. Phys.: Condens. Matter \textbf{19}, 452201 (2007).

\bibitem{Ruff08}
J. P. C. Ruff, J. P. Clancy, A. Bourque, M. A. White, M. Ramazanoglu,
J. S. Gardner, Y. Qiu, J. R. D. Copley, M. B. Johnson, H. A. Dabkowska,
and B. D. Gaulin,
Phys. Rev. Lett. {\bf 101}, 147205 (2008).

\bibitem{Sosin10}
S. S. Sosin, L. A. Prozorova, M. R. Lees, G. Balakrishnan,
and O. A. Petrenko,
Phys. Rev. B \textbf{82}, 094428 (2010).

\bibitem{Dalmas12}
P. Dalmas de R\'eotier, A. Yaouanc, Y. Chapuis, S. H. Curnoe,
B. Grenier, E. Ressouche, C. Marin, J. Lago, C. Baines, and
S. R. Giblin, Phys. Rev. B \textbf{86}, 104424 (2012).

\bibitem{Bonville13}
P. Bonville, S. Petit, I. Mirebeau, J. Robert, E. Lhotel,
and C. Paulsen,
J. Phys.: Condens. Matter \textbf{25},  275601 (2013).

\bibitem{Ross14}
K. A. Ross, Y. Qiu, J. R. D. Copley, H. A. Dabkowska, and B. D. Gaulin,
Phys. Rev. Lett. {\bf 112}, 057201  (2014).

\bibitem{Champion04}
J. D. M. Champion and P. C. W. Holdsworth,
J. Phys.: Condens. Matter \textbf{16}, S665 (2004).

\bibitem{Wong13}
A. W. C. Wong, Z. Hao, and M. J. P. Gingras,
Phys. Rev. B \textbf{88}, 144402 (2013).

\bibitem{McClarty14}
P. A. McClarty, P. Stasiak, and M. J. P. Gingras,
Phys. Rev. B \textbf{89}, 024425 (2014).

\bibitem{Zhitomirsky14}
M. E. Zhitomirsky, P. C. W. Holdsworth, and R. Moessner,
Phys. Rev. B \textbf{89}, 140403(R) (2014).

\bibitem{Ke07}
X. Ke, R. S. Freitas, B. G. Ueland, G. C. Lau, M. L. Dahlberg,
R. J. Cava, R. Moessner, and P. Schiffer,
Phys. Rev. Lett. \textbf{99}, 137203 (2007).

\bibitem{Chang10}
L. J. Chang, Y. Su, Y.-J. Kao, Y. Z. Chou, R. Mittal, H. Schneider,
Th. Br\"uckel, G. Balakrishnan, and M. R. Lees,
Phys. Rev. B \textbf{82}, 172403 (2010).


\bibitem{Palmer00}
S. E. Palmer and J. T. Chalker,
Phys. Rev. B \textbf{62}, 488 (2000).

\bibitem{Long89}
M. W. Long, J. Phys.: Condens. Matter {\bf 1}, 2857 (1989).

\bibitem{Bergman07}
D. L. Bergman, R. Shindou, G. A. Fiete, and L. Balents,
J. Phys.: Condens. Matter {\bf 19}  145204 (2007).

\bibitem{Hoglund07}
K. H. H\"oglund, A. W. Sandvik, and S. Sachdev,
Phys. Rev. Lett. {\bf 98}, 087203 (2007).

\bibitem{Eggert07}
S. Eggert, O. F. Sylju\aa sen, F. Anfuso, and M. Andres,
Phys. Rev. Lett. {\bf 99}, 097204 (2007).

\bibitem{Wollny11}
A. Wollny, L. Fritz, and M. Vojta,
Phys. Rev. Lett. \textbf{107}, 137204 (2011).

\bibitem{Sen11}
A. Sen, K. Damle, and R. Moessner,
Phys. Rev. Lett. \textbf{106}, 127203 (2011).

\bibitem{Wollny12}
A. Wollny, E. C. Andrade, and M. Vojta,
Phys. Rev. Lett. {\bf 109}, 177203 (2012).

\bibitem{Creutz87}
M. Creutz,
Phys. Rev. D {\bf 36}, 515 (1987).

\bibitem{Harris74}
A. B. Harris, J. Phys. C \textbf{7}, 1671 (1974).

\bibitem{Campostrini06}
M. Campostrini, M. Hasenbusch, A. Pelissetto, and E. Vicari,
Phys. Rev. B \textbf{74}, 144506 (2006).

\bibitem{Wessel05}
S. Wessel and I. Milat,
Phys. Rev. B \textbf{71}, 104427 (2005).

\bibitem{Tcher04}
O. Tchernyshyov, H. Yao,  and R. Moessner,
Phys. Rev. B \textbf{69}, 212402 (2004).

\bibitem{Khaliullin05}
G. Khaliullin,
Prog. Theor. Phys. Suppl. \textbf{160}, 155 (2005).

\bibitem{Penc04}
K. Penc, N. Shannon,  and H. Shiba,
Phys. Rev. Lett. \textbf{93}, 197203 (2004).


\bibitem{Onoda10}
S. Onoda and Y. Tanaka, Phys. Rev. Lett. \textbf{105}, 047201 (2010).

\bibitem{Ross11}
K. A. Ross, L. Savary, B. D. Gaulin, and L. Balents,
Phys. Rev. X \textbf{1}, 021002 (2011).








\end{thebibliography}
\end{document}